\documentclass[aps,%preprint,
onecolumn,superscriptaddress,showpacs]{revtex4}
\usepackage{epsf}
\usepackage{graphicx}
\usepackage{amsmath}

\begin{document}
\title{Polymer adsorption on a fractal substrate: numerical study}
%                              AUTHORS revtex4-style
%----------------------------------------------------------------------------
\author{Viktoria Blavatska}
\email[]{E-mail: viktoria@icmp.lviv.ua; blavatska@itp.uni-leipzig.de}
\affiliation{Institut f\"ur Theoretische Physik and Centre for Theoretical Sciences (NTZ),\\ Universit\"at Leipzig, Postfach 100\,920,
D-04009 Leipzig, Germany}
\affiliation{Institute for Condensed
Matter Physics of the National Academy of Sciences of Ukraine,\\
79011 Lviv, Ukraine}
\author{Wolfhard Janke}
\email[]{E-mail: Wolfhard.Janke@itp.uni-leipzig.de}
\affiliation{Institut f\"ur Theoretische Physik and Centre for Theoretical Sciences (NTZ),\\ Universit\"at Leipzig, Postfach 100\,920,
D-04009 Leipzig, Germany}

\begin{abstract}
We study the adsorption of flexible polymer macromolecules on a percolation
cluster, formed by a regular two-dimensional disordered lattice at critical
concentration $p_c$ of attractive sites. The percolation cluster is
characterized by a fractal dimension $d_s^{p_c}=91/49$. The conformational
properties of polymer chains grafted to such a fractal substrate are studied
by means of the pruned-enriched Rosenbluth method (PERM). We find estimates
for the surface crossover exponent governing the scaling of the adsorption energy
in the vicinity of the transition point, $\phi_s^{p_c}=0.425\pm0.009$, and for
the adsorption transition temperature, $T_A^{p_c}=2.64\pm0.02$. As expected,
the adsorption is diminished when the fractal dimension of the substrate is
smaller than that of a plain Euclidean surface. 
The universal size and shape
characteristics of a typical spatial conformation which attains a polymer
chain in the adsorbed state are analyzed as well.

\end{abstract}

\maketitle

\vspace*{-1mm}
\section{Introduction}

The conformational properties of polymer macromolecules in the vicinity of substrates are the subject of 
continuous interest in polymer science, playing an important role both in technology (adhesion, stabilization of colloidal dispersions \cite{Dolan74}) and biological physics (proteins adsorption on membranes \cite{Maier99,Xie02}). The presence of an energetically neutral surface produces only 
trivial effects of steric restrictions for polymers. More interesting is the case of an attractive substrate, when below a
critical temperature $T_A$ a second-order phase transition into an adsorbed state takes place  \cite{Eisenriegler93}. 
The peculiarities of adsorption of grafted polymers on attractive surfaces 
are thoroughly studied by now both analytically \cite{Diehl81, Eisenriegler82, Gennes83, Diehl94, Usatenko06} 
and numerically \cite{Eisenriegler82, Meirovitch88,Hegger94,Descas04,Grassberger05,Janke1,Janke2}.  
As an order parameter of the adsorption transition, one considers the fraction of the average number of monomers $N_s$ adsorbed to the 
surface and the total length $N$ of the polymer chain, which tends to zero in the usual bulk regime and becomes macroscopic close
 to $T_A$, obeying the scaling law
 \begin{equation}
 \frac{\langle N_s \rangle}{N}\sim N^{\phi_s-1},\,\,\,\,\,\,N\to\infty.
 \end{equation}    
Here, $\phi_s$ is the surface crossover exponent, a basic parameter
in scaling analysis of the adsorption transition ($0< \phi_s < 1$).  
Recent estimates of the crossover exponent $\phi_s$ along with numerical values for the adsorption temperature $T_A$ are given in Table 1. 

\begin{table}[b!]
\caption{ Crossover exponent $\phi_s$ and adsorption critical temperature $T_A$ for polymers grafted on a homogeneously attractive plain surface 
and on the fractal surface formed by a percolation cluster ($\phi_s^{p_c}$, $T_A^{p_c}$). RG: renormalization group studies, MC: Monte Carlo simulations.}
\label{dim}
\begin{center}
\begin{tabular}{lcccc}
\hline
\hline
  & $\phi_s$ & $T_A$ & $\phi_s^{p_c}$ & $T_A^{p_c}$  \\
\hline
RG & 
     0.482 \cite{Diehl81} & &  & \\
  &   0.518 \cite{Diehl94}  & &  & \\
MC  
%&  $0.530\pm0.007$ \cite{Meirovitch88} & $3.436\pm0.01$\cite{Meirovitch88} &    \\
%& 0.59 \cite{Descas04} & & &\\
   &     $0.496\pm0.005$ \cite{Hegger94}  & $3.497\pm0.003$ \cite{Hegger94} & $0.425\pm0.009$ (this study) & $2.64\pm0.02$ (this study) \\
    &   $0.484\pm0.002$ \cite{Grassberger05} &  $3.5006\pm0.0009$ \cite{Grassberger05} &    &  \\
\hline\hline
\end{tabular}
\end{center}
\end{table}

The study of polymers near disordered surfaces is of great importance, since most
naturally occurring substrates are rough and energetically (or structurally) inhomogeneous. 
Surface heterogeneity has a crucial effect on polymer adsorption phenomena \cite{Baum90,Balazs91,Kawaguchi91,Baumgartner91,Huber98,Sumithra98,Moghaddam03,Usatenko07,Ziebarth07}. In fact, 
already simple physical arguments lead to the conclusion that upon increasing the surface irregularity 
the number of polymer-surface contacts is strongly influenced, leading to a shift of the adsorption critical temperature.
Energetical inhomogeneity arises due to the presence of various chemical compounds in the substrate, interacting 
with the monomers of the polymer chain in a different manner. In the language of lattice models, such surfaces can be 
modeled as a two-dimensional regular lattice with
different types of randomly distributed sites, e.g., one type with attractive interactions with the monomers 
 and the other one being neutral (treated as defects or impurities).    
Similarly as it holds in the bulk case \cite{Kim87,Nakanishi92}, presence of uncorrelated point-like defects of low concentration (well below the percolation 
threshold $p_c=0.592746$ {\cite{Ziff94}) is expected to be irrelevant for the scaling properties of the adsorption transition of polymers.
Numerical simulations \cite{Sumithra98,Ziebarth07} reveal, however, a continuous dependence of the transition temperature $T_A$ on the  
concentration $p$ of attractive sites. In particular, close to $p_c$ the estimate $T_A^{p_c}\simeq 2.3$ was obtained. 
% was found: 
%$ \phi_s(p)\approx 0.35+0.2 p$.  
The related problem of the impact of long-ranged correlations in the distribution of defects on the surface, leading to a non-trivial influence on 
scaling near the adsorption transition point, was studied recently in Ref.~\cite{Usatenko07}.

Since most chemical substrates are proved to be of fractal nature \cite{Avnir83}, studying the 
influence of a non-trivial surface geometry on polymer adsorption is of particular interest. In 
Ref.~\cite{Bouchaud89} it was found, that the crossover exponent $\phi_s$ 
for a substrate characterized by the fractal dimension $d_s^f$, has upper and lower bounds given as
\begin{equation}
1-(3-d_s^f)\nu \leq  \phi_s \leq d_s^f/3,
\end{equation}
where  $\nu$ is the bulk radius of gyration exponent for a polymer 
chain in a good solvent ($\nu=0.5887\pm0.0006$ \cite{Donald92}). One can thus conclude that adsorption is enhanced (diminished) when the fractal dimension of the substrate is larger (smaller) than that of a plain Euclidean surface. A number of studies has been dedicated to polymer adsorption on a family of finitely ramified fractals \cite{Elesovic02,Bubahja93,Kumar93,Miljkovic95}.  
Also of great importance is the study of polymers in the vicinity of fluctuating surfaces, such as membranes \cite{Auth03,Karalus11}.

\begin{figure}[t!]\begin{center}
\includegraphics[width=5cm]{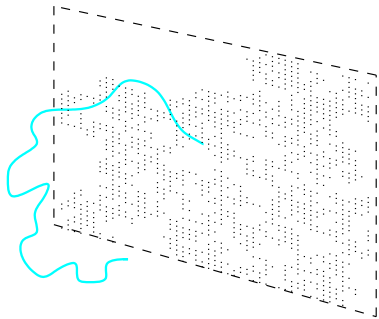}
\caption{ \label{mem} Sketch of a polymer chain grafted to an attractive ``sieve'' formed by a percolation cluster.}\end{center}
\end{figure}

In this concern, it is worthwhile to study the situation when the concentration of attractive sites 
on the surface is exactly at the percolation threshold and a spanning percolation cluster
of attractive sites appears. A percolation cluster is a fractal object with fractal dimension $d_s^{p_c}=91/49\simeq1.89$ {\cite{Havlin87}}.
In general, studying polymer adsorption on a percolative surface, one encounters two possible statistical averages. 
In the  first (considered previously in Refs.~\cite{Sumithra98,Ziebarth07}), the statistical ensemble includes all attractive sites on the surface, whereas 
in the second, one takes into account only sites belonging to the percolation cluster.  
In the present study, we consider the particular situation, when the neutral sites of the surface (which do not belong to the percolation cluster)  
  are penetrable for the polymer chain, and the polymer is adsorbed on the attractive fractal with fractal dimension  $d_s^{p_c}$. 
  This can model the process of polymer adsorption on an attractive, 
  partially penetrable ``sieve'' (see Fig.~{\ref{mem}}), which could be of interest in biophysical applications.

%\begin{table}[t!]
%\small{
%\caption{ Values of $\Theta$-temperature on pure regular lattices and on the site-diluted lattices at the percolation threshold for different space dimensions $d$. SE: series enumerations, MC: Monte Carlo simulations}
%\label{dim}
%\begin{center}
%\begin{tabular}{lcr}
%\hline  & $d=2$ & $d=3$ \\
%\hline
%$\Theta_{pure}$ SE  &  1.520 \cite{Foster92}  & \\
%  MC  & 1.497 \cite{Barkema98} & 2.972$\pm 0.006$ \cite{Szleifer92} \\
%MC & & 3.717$\pm0.003$ \cite{Grassberger97}\\
%Eq. (\ref{tpure}) &  & \\
%\hline
% $\Theta_{p_c}$  SE & 0.71$\pm0.08$ \cite{Barat92} & 0.43$\pm0.06$ \cite {Barat95} \\
% & 0.67$\pm0.06$ \cite{Barat91} &\\
%Eq. (\ref{tpc}) & 0.88 &1.158 \\
% \hline
%\end{tabular}
%\end{center}
%}
%\end{table}
\section{The method}

We start with a regular two-dimensional lattice of edge lengths up to $L_{{\rm max}}{=}400$, each site of 
which is assigned to be occupied with percolation probability $p_c$ and empty otherwise. 
To extract the percolation cluster of occupied sites, which spans around the lattice, an algorithm 
based on the  site-labeling method of Hoshen and Kopelman \cite{Hoshen76} has been applied. Note, that the definition of spanning clusters 
on finite lattices is not unique, in particular one could consider clusters connecting only two opposite borders.  
 In this case, however, the constructed clusters are anisotropic in space and could be related to the problem of so-called directed percolation \cite{Family82}.
% which are characterized by the same fractal dimension and are equally legitimate. The definition which we 
We therefore take only incipient clusters into account which reach the borders of the lattice in {\em all\/} coordinate directions and hence are expected to be
 more isotropic. 

The polymer chain is modeled as a self-avoiding walk (SAW).
To study the conformational properties of SAWs, grafted to the substrate formed by a percolation cluster,  we apply the pruned-enriched Rosenbluth method (PERM) \cite{Grassberger97}.
The starting point of a SAW is fixed on a
  random site which belongs to the percolation cluster (see Fig.~\ref{mem}). 
 Note, that this starting site is always chosen within a small region around the center of a given percolation cluster to allow the
adsorbed polymer chain configurations to be completely located on the cluster.
We treat this disordered surface as 
the $z=0$ plane of a regular three-dimensional lattice. 
The chain grows step by step, i.e., the $n$th monomer is placed at a randomly chosen neighbor site of the last placed $(n-1)$th  monomer ($n\leq N$), taking into account that the chain cannot ``penetrate'' through the occupied sites of the surface (belonging to the percolation cluster), but only through the empty sites. The growth is stopped, 
if the total length of the chain, $N$, is reached (we consider SAWs of length up to $N=150$).
The adsorption energy $E_n$ of a growing chain at the $n$th step is given by
\begin{equation}
E_n=N_s(n)\,\varepsilon, \label{en}
\end{equation}
where  $\varepsilon$ is the attractive energy between monomers and the percolation cluster sites  
and $N_s(n)$ is the number of contacts of the polymer chain with attractive sites.  

%To obtain the correct statistics, any attempt to place a new monomer at a site which already contains a monomer would result in
% discarding the entire chain. This leads to an exponential ``attrition" (the number of discarded chains grows exponentially with the chain length). 
%The bias due to avoiding this case is corrected by means of giving 
A weight $W_n$ is given to each sample configuration at the $n$th step, which in our case is given by
\begin{equation}
W_n=\prod_{l=2}^n m_l {\rm e}^{-\frac{E_l-E_{l-1}}{k_B T}}.
\label{weight}
\end{equation} 
Here,  $m_l$ is the number of free lattice sites to place the $l$th monomer and $k_B$ is the Boltzmann constant.
In what follows, we will assume units in which $\varepsilon=-1, k_B=1$.
%Population control in 
%PERM suggests pruning configurations with too small weights, and enriching the sample with copies of high-weight 
%configurations \cite{Grassberger97}. These copies are made while the chain is growing, and continue to grow independently of each other. 
Pruning and enrichment are 
performed by choosing thresholds $W_n^{<}$
and $W_n^{>}$ depending on the current estimate of the sum of weights  $Z_n=\sum_{{\rm conf}} W_n^{{\rm conf}}$ of the $n$-monomer chain \cite{Grassberger97,Hsu03,Bachmann03}. 
If the current weight $W_n$ of an $n$-monomer chain is less than $W_n^{<}$,  the chain is discarded with probability 
$1/2$, whereas if $W_n$ exceeds  $W_n^{>}$, the configuration is doubled (enrichment of the sample with high-weight configurations). 
%For updating the threshold values we apply similar rules as in : $W_n^{>}{=}C(Z_n/Z_1)(c_n/c_1)^2$ and $W_n^{<}{=}0.2W_n^{>}$, where $c_n$ denotes the number of %created chains having length $n$, and the parameter $C$ controls the pruning-enrichment statistics. 
%After a certain number of chains of total length $N$ is produced, the iteration is finished and a new tour starts. 
%We adjust the pruning-enrichment control parameter such that on average 10 chains of total length $N$ are generated per each iteration \cite{Bachmann03}, 
%and perform $10^5$ iterations. 
%When a chain of total length $N$ is constructed, a new one starts, until the desired number of chain configurations is obtained.  

The configurational averaging for any observable $O$ is given by
\begin{eqnarray}
&&\langle O \rangle  
=\frac{\sum_{{\rm conf}} W_N^{{\rm conf}}O } {\sum_{{\rm conf}} W_N^{{\rm conf}}}, \label{R}
%&&P(R,N){=}W_N^{{\rm conf}}\sum_{{\rm conf}} W_N \label {Z},
\end{eqnarray}
where  $W_N^{{\rm conf}}$ is the weight of an $N$-monomer chain in a given configuration.
In the problem under consideration, a double averaging has to be performed: The first $\langle ...\rangle$  over all 
configurations of the polymer chain grafted to a single percolation cluster;
the second average $\overline{ \langle ... \rangle}$ is carried out over different realizations of disorder, i.e., over 
different constructed percolation clusters:
\begin{eqnarray}
&&\overline{\langle O \rangle}=\frac{1}{M}\sum_{i{=}1}^M \langle O\rangle_i. \label{avprob} 
\end{eqnarray}
Here, $M$ is the number of different clusters and the index $i$ means that a given quantity is calculated on cluster $i$. 
We constructed $M=1000$ clusters. Note, that the case of so-called ``quenched disorder'' is considered, where 
the average over different disorder realizations is taken after the configurational average has been performed.

\section {Results}

The adsorption transition is in general viewed as a second-order phase transition \cite{Eisenriegler93} with the averaged fraction of monomers 
on the surface $\langle N_s\rangle/N$ viewed as order parameter. Note that this value can also be interpreted as an adsorption 
energy per monomer (cf.\ Eq.~(\ref{en})). In the thermodynamic limit $N\to\infty$, the adsorption energy tends to zero in the desorbed phase for $T>T_A$ 
and becomes macroscopic close to the
transition point, where it scales according to (1) \cite{Eisenriegler82}:
\begin{equation}
  \langle N_s \rangle /N \sim \left\{
  \begin{array}{ll}
  \frac{1}{(T-T_A)N},& \mbox{ $T>T_A$},\\
  N^{\phi_s-1},& \mbox{ $T=T_A$},\\
(T_A-T)^{\frac{1-\phi_s}{\phi_s}}, & \mbox{ $T<T_A$} .
  \end{array}\right.\label{}
  \end{equation}
In the adsorbed phase for $T<T_A$, the fraction $\langle N_s \rangle/N$ is independent of $N$. Introducing the scaling 
variable $x = |T-T_A| N^{\phi_s}$, the adsorption energy per monomer can be presented in general in
the scaling form 
 \begin{equation}
\langle N_s\rangle/N = N^{\phi_s-1}F(|T-T_A| N^{\phi_s})\label{sc}
\end{equation}
 with
\begin{equation}
 F(x) \sim \left\{
  \begin{array}{ll}
  \frac{1}{x},& \mbox{ $T>T_A$},\\
  {\rm const},& \mbox{ $T=T_A$},\\
x^{\frac{1-\phi_s}{\phi_s}}, & \mbox{ $T<T_A$} .
  \end{array}\right.\label{g}
  \end{equation}

Our analysis of the temperature behavior of the order parameter $\overline{\langle N_s \rangle}/N$ for chain lengths up to $N=140$ is shown in Fig.~\ref{Ns}
(for comparison and to check the validity of our computer code, we re-consider the case of a homogeneous attractive surface as well). 
The number of contacts with attractive sites of the surface increases monotonically as the temperature is lowered  
and becomes macroscopic within a short temperature interval
close to the adsorption transition.  Whereas for the case of a homogeneously attractive surface $\overline{\langle N_s\rangle}/N$ reaches its maximum value 
close to $1$ at $T\ll T_A$ as expected (the polymer lies on the $z=0$ plane), in the case of a fractal surface this value is 
 found to be slightly smaller. Due to the complicated structure of a percolation cluster  (in particular the existence of 
numerous ``dead-ends") even at very low temperatures some small percentage of 
monomers occupy sites of the $z=0$ plane which do not belong to the cluster (as we checked explicitly for idealized clusters constructed by hand) and according to
our definition (\ref{en}) are {\em not\/} counted as ``adsorbing''; the ground state with lowest energy is thus not reached. This is a dynamic problem which is also encountered in other disordered systems, e.g. in spin glasses. It requires a very long observation time
for a polymer to find a configuration  completely located on the edges of the percolation cluster. 
%If we do count all
%monomers in the $z=0$ plane, the limiting value $1$ for $T \rightarrow 0$ is restored, i.e., for
%very low temperatures also for the attractive percolation cluster the polymer conformations are 
%strictly two-dimensional. 

\begin{figure}[t!]\begin{center}
\includegraphics[width=7cm]{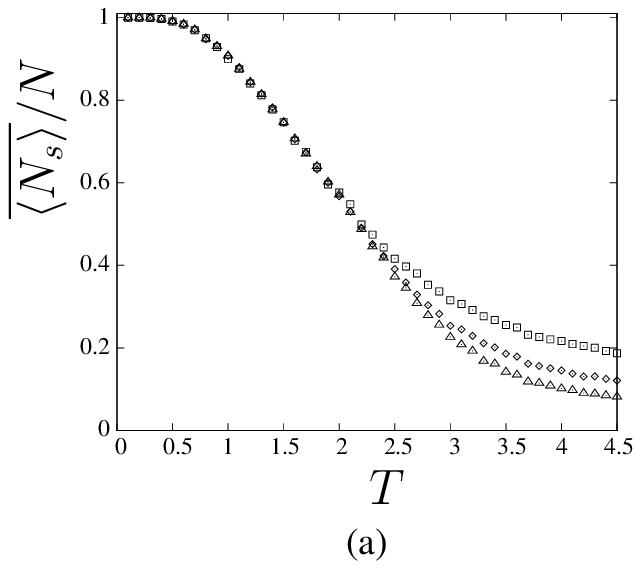}
\hspace*{1cm}
\includegraphics[width=7.1cm]{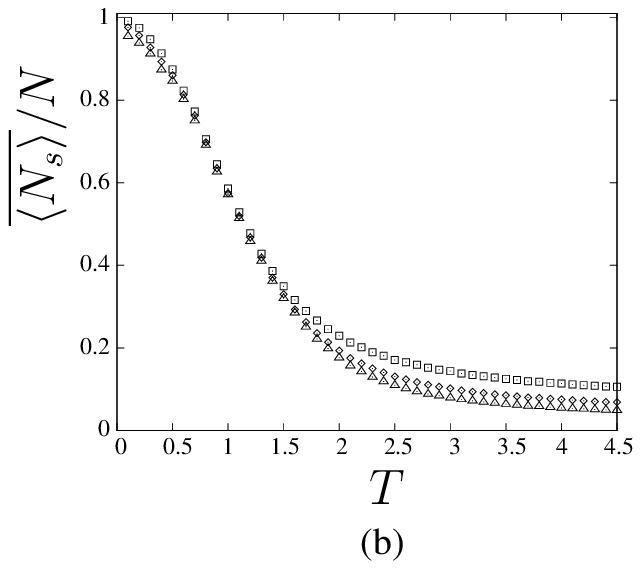}
\caption{ \label{Ns} Averaged fraction of monomers of the chain adsorbed on (a) a homogeneously attractive surface
 and (b) an attractive
 percolation cluster  for $N=40$ (squares) $N=80$ (diamonds), $N=140$ (triangles) as a function of temperature. 
%(a) and as a function of $N$ at $T=0.2$ (b).
} 
\end{center}
\end{figure}

\begin{figure}[b!]
\begin{center}
\includegraphics[width=8.5cm]{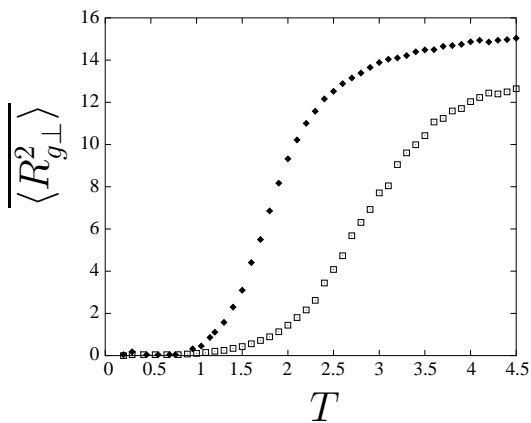}
\caption{ \label{Rgperp} Component of gyration radius of the polymer chain in direction perpendicular to 
the surface  for the cases of a homogeneously attractive surface
 (squares) and an attractive 
 percolation cluster (filled diamonds) for $N=140$ as a function of temperature.
}. 
\end{center}
\end{figure}

Due to the presence of the surface, which breaks the space isotropy, one distinguishes between the polymer size characteristics in directions
parallel and perpendicular to the surface.
Let $\vec{R}_n=\{x_n,y_n,z_n\}$ be the position vector of the $n$th monomer of the polymer chain ($n=1,\ldots,N$).
The components of squared radius of gyration in direction parallel and perpendicular to the surface are given by:
\begin{equation}
R_{g||}^2 =\frac{1}{2N^2} \sum_{n=1}^N\sum_{m=1}^N \left[(x_n-x_m)^2+(y_n-y_m)^2\right],\,\,\,\,\,\,\,
R_{g\perp}^2 =\frac{1}{2N^2} \sum_{n=1}^N\sum_{m=1}^N (z_n-z_m)^2.
\label{Rgs}
\end{equation}

The component of the gyration radius in direction perpendicular to the surface ($z$-direction), ${\langle R^2_{g\perp}\rangle}$, 
  can be interpreted as the average thickness of the layer of adsorbed monomers.  Well above the transition temperature, 
  it obeys the usual bulk scaling behavior and becomes $N$-independent in the adsorbed phase \cite{Eisenriegler82}:
\begin{equation}
\langle R^2_{g{\perp}}\rangle \sim\left\{
 \begin{array}{ll}
N^{2\nu} ,& \mbox{ $T>T_A$},\\
(T_A-T)^{-\frac{2\nu}{\phi_s}}, & \mbox{ $T<T_A$}.
\end{array}\right.
\end{equation} 
Here, $\nu$ is a well-known universal critical exponent, governing the scaling of the radius of gyration in the bulk 
($\nu=0.5887\pm0.0006$ \cite{Donald92}).
The corresponding scaling ansatz is then
\begin{equation}
\langle R^2_{g{\perp}} \rangle \sim N^{2\nu}G_{\perp}(|T-T_A| N^{\phi_s}), \label{perp}
\end{equation}
 with scaling function $G_{\perp}(x)={\rm const}$ for $T>T_A$ and $G_{\perp}(x)=x^{-\frac{2\nu}{\phi_s}}$ for $T<T_A$.

The component of the gyration radius in direction parallel to the surface, $\langle R^2_{g{||}} \rangle $, has similar scaling 
behavior: For $T>T_A$ the usual bulk behavior is reproduced, whereas for $T<T_A$ the polymer chain predominantly lies on the surface 
and behaves statistically as two-dimensional \cite{Eisenriegler82}:
\begin{equation}
\langle R^2_{g{||}} \rangle \sim\left\{
 \begin{array}{ll}
N^{2\nu} ,& \mbox{ $T>T_A$},\\
N^{2\nu_2}(T_A-T)^{-\frac{2\nu_2-\nu}{\phi_s}}, & \mbox{ $T<T_A$}.
\end{array}\right.
\end{equation} 
where $\nu_2$ is the corresponding critical exponent in two dimensions
($\nu_2=3/4$ \cite{Nienhuis82} in the homogeneous case). Again, this allows a scaling representation: 
\begin{equation} \langle R^2_{g{||}}\rangle \sim N^{2\nu}G_{||}(|T-T_A| N^{\phi_s}), \label{par}\end{equation} with scaling function 
$G_{||}(x)={\rm 
  const}$ for $T>T_A$ and
  $G_{||}(x)=x^{\frac{2\nu_2-\nu}{\phi_s}}$ for $T<T_A$.

\begin{figure}[h!]
\begin{center}
\includegraphics[width=8.5cm]{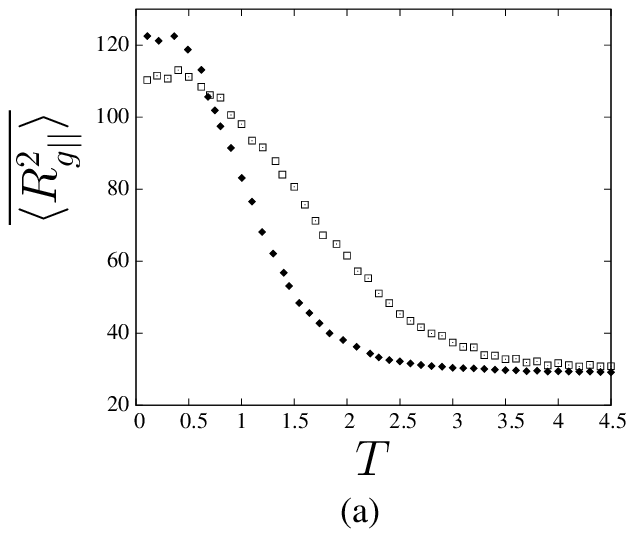}
\hspace*{0.2cm}
\includegraphics[width=8.5cm]{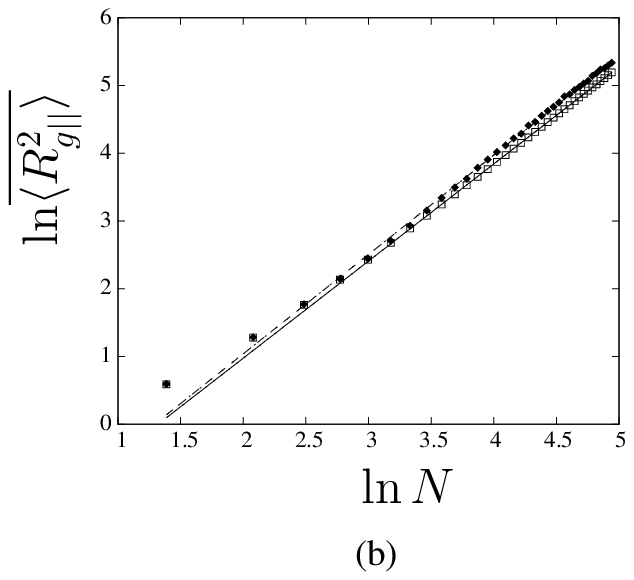}
\caption{ \label{Rgpar} Component of gyration radius of the polymer chain in direction parallel to 
the surface for the cases of a homogeneously attractive surface
 (squares) and an attractive 
 percolation cluster (filled diamonds) as (a) a function of temperature and (b) as a function of $N$ in double logarithmic scale at $T=0.1$. Solid line: least-square fitting with $\nu_2=0.742\pm0.006$, dashed line: least-square fitting with $\nu_2^{p_c}=0.772\pm0.006$.}
\end{center}
\end{figure}

Our results for $\overline{\langle R^2_{g{\perp}}\rangle}$ as a function of temperature are presented in Fig.~\ref{Rgperp}.  At each temperature, 
the polymer layer thickness on the homogeneous surface is smaller than that on the percolation cluster 
 due to stronger attraction to the surface. For $T < 0.5$ the layer thickness is so small that the conformations are basically two-dimensional in both cases. $\overline{\langle R^2_{g||}\rangle}$ as function of temperature is shown in Fig.~\ref{Rgpar}(a).  Examining the $N$-dependence 
of the parallel component of the gyration radius at temperatures well below the 
adsorption point [Fig.~\ref{Rgpar}(b)], we can find estimates of the critical exponent $\nu_2$ by least-square fitting. For the case of a homogeneous surface, the value 
 $\nu_2=0.742\pm0.006$ is restored. 
 For the critical exponent, governing the scaling for a polymer chain adsorbed on an attractive percolation cluster, 
  a  value $\nu_2^{p_c}=0.772\pm0.006$ is obtained. This exponent is compatible  with the  one for the average size of a polymer residing on the the 
sites of a two-dimensional percolating 
cluster,  $\nu_2^{p_c}=0.782\pm0.003$ \cite{Blavatska08}.

\begin{figure}[b!]\begin{center}
\includegraphics[width=6.7cm]{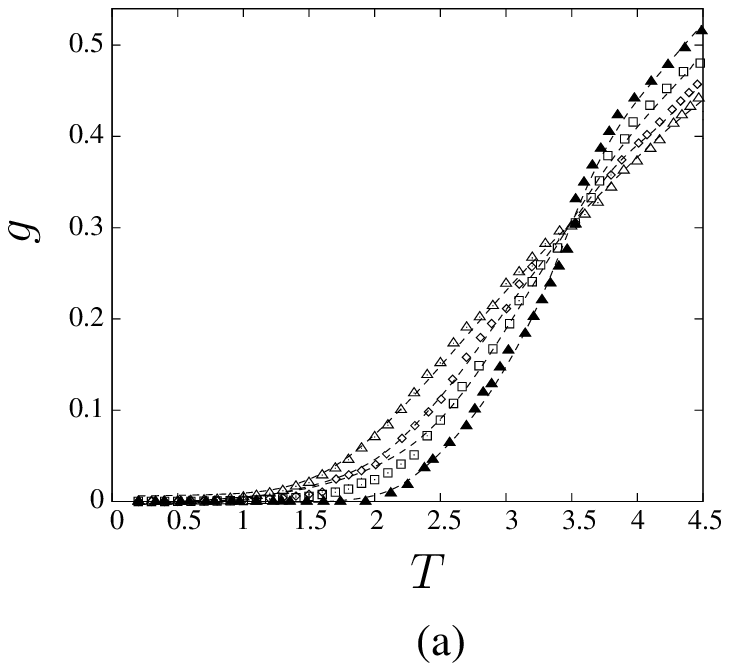}
\hspace*{1cm}
\includegraphics[width=6.7cm]{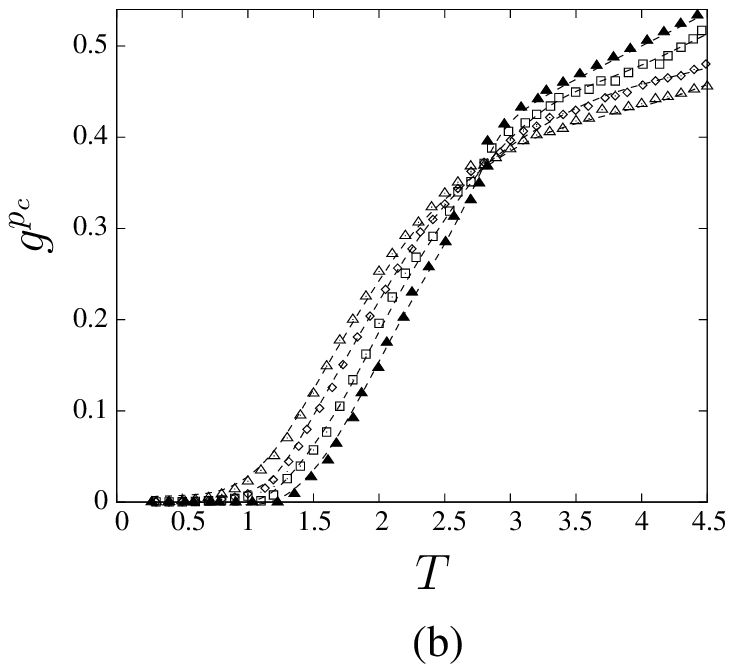}
\caption{ \label{ratio} The size ratio  $g=\overline{\langle R^2_{G\perp}\rangle}/\overline{\langle R^2_{g{||}}\rangle}$ of the polymer chain near (a) a homogeneously attractive surface and (b) an attractive 
 percolation cluster as a function of temperature. Triangles: $N=80$, diamonds: $N=100$, squares: $N=120$, filled triangles: $N=140$. }\end{center}
\end{figure}

The study of the size ratio  $g\equiv \overline{\langle R^2_{g\perp} \rangle}/\overline{\langle R^2_{g{||}} \rangle}$ can be used to 
estimate the critical adsorption temperature.
Remembering the scaling representations  of the components of the gyration radius (\ref{perp}) and (\ref{par}), 
one has: 
\begin{equation}
g = G_{\perp}(|T-T_A| N^{\phi_s})/G_{||}(|T-T_A| N^{\phi_s}) = G(|T-T_A| N^{\phi_s}).
\end{equation}
At the adsorption critical point ($T\to T_A$), this ratio  
becomes independent of $N$ and thus, when plotting $g$ vs $T$ for different $N$, all curves should intersect in a single point which namely gives $T_A$.
In Fig. \ref{ratio}, we present our results for the size ratio in the cases of (a) a homogeneous attractive surface and (b) the fractal substrate.
The range of positions of points of intersection enables us to obtain estimates of the adsorption transition critical temperature:
$T_A=3.5\pm0.1$, $T_A^{p_c}=2.7\pm0.1$.

\begin{figure}[b!]\begin{center}
\includegraphics[width=8.5cm]{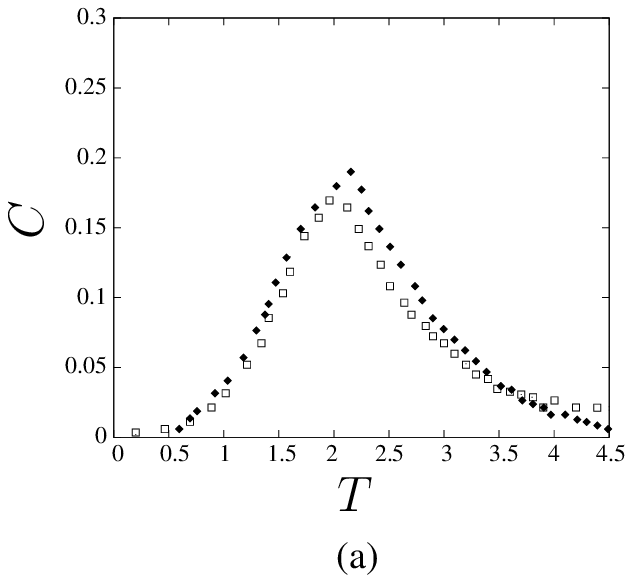}
\hspace*{0.2cm}
\includegraphics[width=8.9cm]{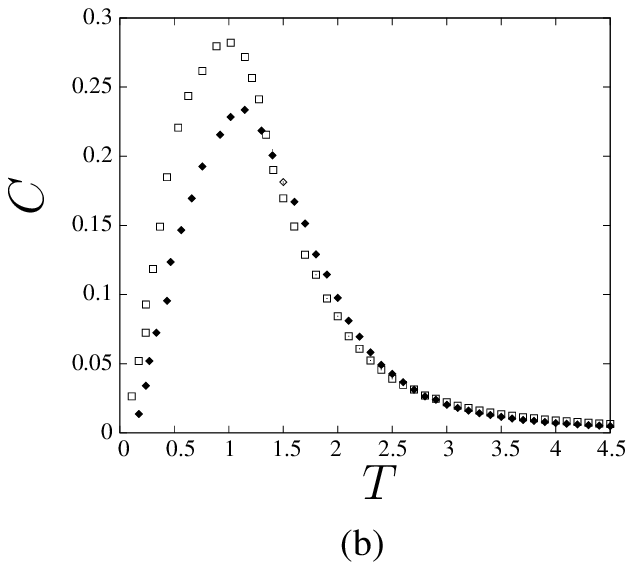}
\caption{ \label{Cv} Specific heat per monomer as a function of temperature for a polymer chain near (a) a homogeneously attractive surface
 and (b) an attractive 
 percolation cluster as a function of temperature. Squares: $N=40$, filled diamonds: $N=100$.}\end{center}
\end{figure}

The characteristics of the adsorption transition can be obtained by examining the fluctuations of the adsorption energy near the transition point. The specific heat per monomer is given by
 \begin{equation}
C=\frac{1}{NT^2}\left(\overline{\langle N_s^2 \rangle}- \overline{\langle N_s \rangle^2}\right).
\label{cvperc}
\end{equation}
Taking into account Eq.~(\ref{sc}), one obtains the scaling form for the specific heat \cite{Eisenriegler82}:
\begin{equation}
C\sim N^{2\phi_s-1}H(|T-T_A| N^{\phi_s}). \label{cvsc}
\end{equation}   
The peak structure of $C$ as a function of temperature indicates transitions or crossovers between physically different states. In the problem under consideration,
this corresponds to the transition between bulk and adsorbed regimes. Figure \ref{Cv} shows the typical specific-heat behavior of SAWs grafted to a homogeneous surface 
and percolation clusters, respectively.
\begin{figure}[t!]\begin{center}
%\psfrag{Cmax}{{\Large${C_{{\rm max}}}(N)$}}
\includegraphics[width=8.9cm]{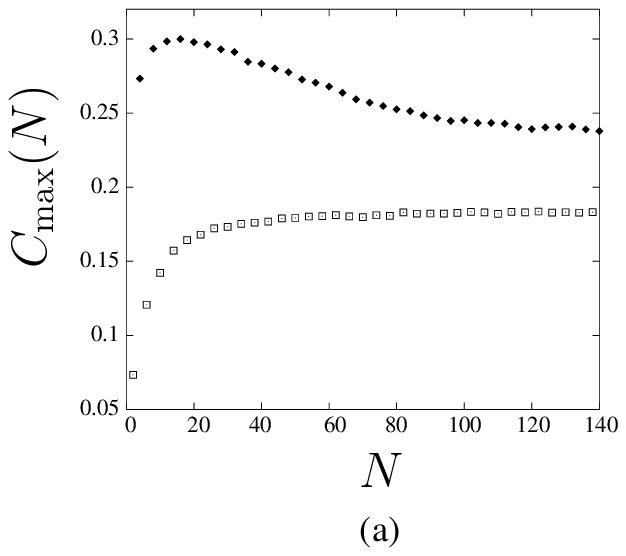}
\hspace*{0.2cm}
\includegraphics[width=8.5cm]{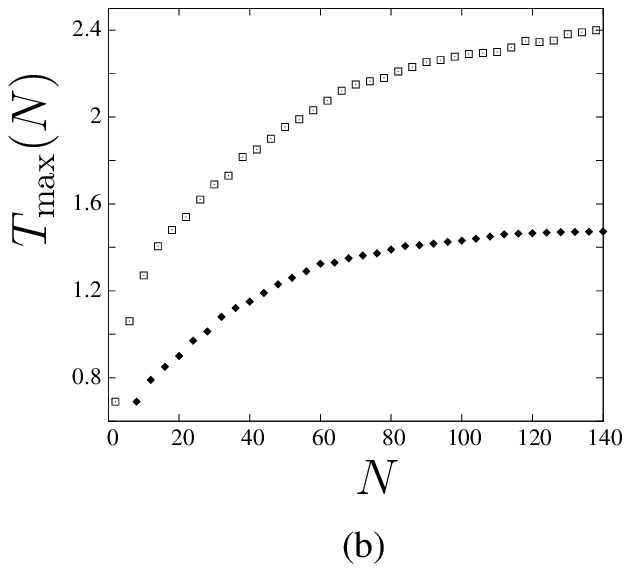}
\caption{ \label{Cvmax} (a) The maximum heights of the specific-heat curves and (b) the temperatures defined by the position of the specific-heat maximum 
for a polymer chain near a homogeneously attractive surface
 (squares) and an attractive  percolation cluster (filled diamonds) as functions of $N$. }\end{center}
\end{figure}
Note, that the maximum of the specific heat per monomer grows with 
$N$ for the case of a plain surface, whereas for the case of a fractal surface it decreases with increasing $N$.
Assuming that the value of the specific heat at its maximum (the height of the specific-heat curve) $C_{{\rm max}}(N)$ at each $N$ is already close enough to the 
asymptotic region where 
(\ref{cvsc}) holds, we can estimate the crossover exponent $\phi_s$ by fitting the curves in Fig.~\ref{Cvmax}(a) to the form
\begin{equation}
C_{{\rm max}}(N)\sim a+bN^{2\phi_s-1},
\end{equation}
where $a$ and $b$ are some constants. We obtain $\phi_s=0.509\pm0.009$, $\phi_s^{p_c}=0.425\pm0.009$. 

\begin{figure}[b!]\begin{center}
\includegraphics[width=7.3cm]{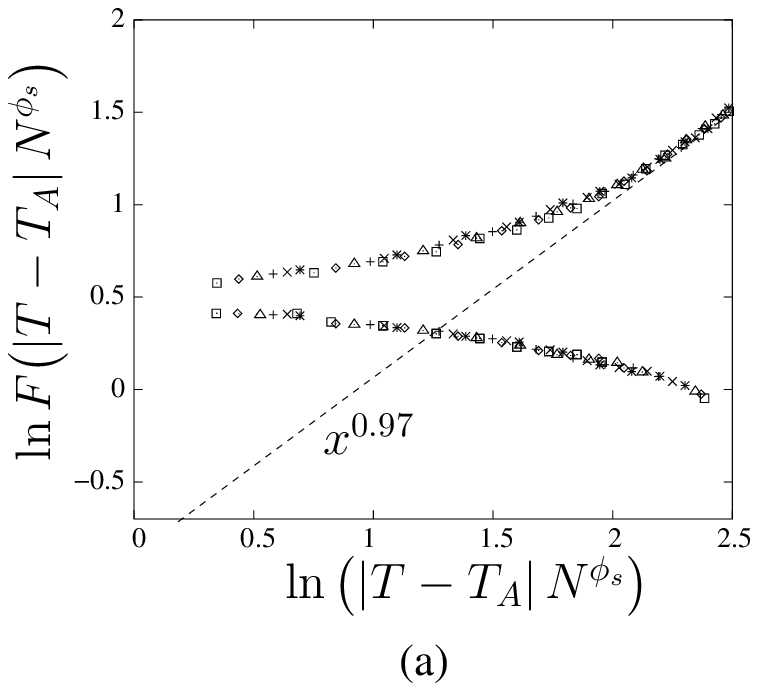}
\hspace*{1cm}
\includegraphics[width=7.3cm]{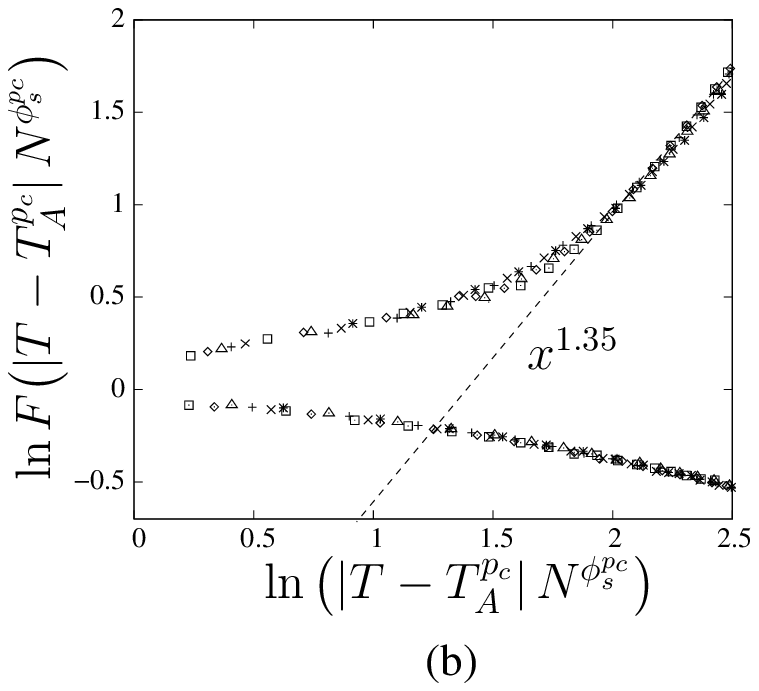}
\caption{ \label{Nsscal} The scaling function (\ref{sc}) as a function of its argument for a polymer chain near (a) a homogeneously attractive surface
and (b) an attractive
 percolation cluster. Dashed lines are results of fitting to the scaling form (\ref{g}) for $T<T_A$.  Squares: $N=40$, diamonds: $N=60$, triangles: $N=80$, pluses: $N=100$, crosses: $N=120$, stars: $N=140$.} 
\end{center}
\end{figure}

For finite chain length $N$, the temperature defined by the position of the specific-heat maximum $T_{{\rm max}}(N)$ is well below 
the transition temperature $T_A$ of an infinitely long polymer macromolecule. This finite-size deviation obeys a scaling behavior:
\begin{equation}
T_{\rm max}(N)-T_{A}\sim N^{-\phi_s}.
%\,\,\,\,\,\,\,\phi=\frac{\nu_{\Theta}}{\nu_p},
\label{tcv}
\end{equation}
Fitting the curves in Fig.~\ref{Cvmax}(b) to this form, and making use of the estimates for $\phi_s$ found by us,
we receive for the critical temperature of the adsorption transition onto a homogeneous surface $T_A=3.47\pm0.02$, and for the case of a percolation cluster the result of fitting gives $T_A^{p_c}=2.64\pm0.02$.

The values obtained could be verified by plotting, e.g., the scaling function of the order parameter (\ref{sc}) as a function of its argument in double logarithmic scale for different chain 
lengths $N$ (Fig.~\ref{Nsscal}). As expected, a data collapse is obtained.  
 The ``upper'' branches in both plots correspond to $T<T_A$ and  
scale  with their argument as $x^{\frac{1-\phi_s}{\phi_s}}$ according to (\ref{g}). The ''lower'' branches
corresponding to $T>T_A$, after reaching the asymptotic limit, should decrease according to (\ref{g}) as $x^{-1}$.

Finally, we analyze the shape of a typical spatial conformation, which attains a polymer chain in the adsorbed state. The measure of the shape properties of a polymer chain in $d$ dimensions can be characterized \cite{Solc71,Rudnick86} in terms of the gyration tensor $\bf{Q}$ with components
\begin{equation}
Q_{ij}=\frac{1}{2N^2}\sum_{n=1}^N\sum_{m=1}^N(x_n^i-{x^i_m})(x_n^j-{x^j_m}),\,\,\,\,\,\,i,j=1,\ldots,d,
\label{mom}
\end{equation}
where $x^i_n$ are the components of the position vector ${\vec{R}_{n}}$. 
Special cases are the squared radius of gyration parallel and perpendicular to the surface in 
(\ref{Rgs}), $R_{g||}^2 = Q_{11}+Q_{22}$ and $R_{g\perp}^2=Q_{33}$. In general,
the spread in eigenvalues $\lambda_i$ of the gyration tensor describes the distribution of monomers inside the polymer coil and
thus measures the asymmetry of a molecule; in particular, for a symmetric (spherical)
configuration all the eigenvalues $\lambda_{i}$ are equal, whereas for completely stretched, rod-like configurations 
all eigenvalues are zero except of one. To compute the quantities $\lambda_i$ analytically is, however, 
difficult, because one must explicitly diagonalize the gyration tensor for each realization in an ensemble of polymers. 
It was thus proposed \cite{Aronovitz86,Rudnick86} to characterize the asymmetry of polymer configurations by rotationally invariant universal quantities, 
such as the asphericity $A_d$, defined as:
\begin{equation}
{A_d} =\frac{1}{d(d-1)} \sum_{i=1}^d\frac{(\lambda_{i}-{\overline{\lambda}})^2}{\overline{\lambda}^2}=
\frac{d}{d-1}\frac{\rm {Tr}\,\bf{\hat{Q}}^2}{(\rm{Tr}\,{\bf{Q}})^2}, \label{add}
\end{equation}
with $\overline{\lambda}=(1/d)\sum_{i=1}^d\lambda_i$ and ${\bf{{\hat{Q}}}}\equiv{\bf{Q}}-\overline{\lambda}\,{\bf{I}}$ (here $\bf{I}$ is  the unity matrix).
This universal quantity equals zero for a spherical configuration and takes a maximum value 
of one in the case of a rod-like configuration. Thus, the inequality $0\leq A_d\leq 1$ holds.
Numerous studies indicate that a  typical flexible polymer chain in good (bulk) solvent takes on the shape of an elongated ellipsoid
with $\langle A_2 \rangle=0.501\pm 0.003$ \cite{Bishop88}, $\langle A_3 \rangle=0.431\pm0.002$ \cite{Jagodzinski92}.

\begin{figure}[t!]\begin{center}
%\psfrag{T}{{\Large${T}$}}
%\psfrag{A}{{\Large$\overline{\langle A_3 \rangle}$}}
\includegraphics[width=8.0cm]{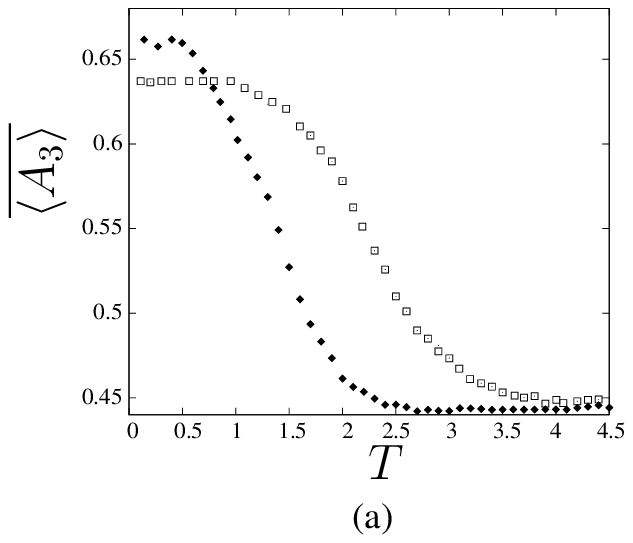}
\hspace*{0.2cm}
%\psfrag{N}{{\Large${N}$}}
%\psfrag{AA}{{\Large$\overline{\langle A_2 \rangle}$}}
\includegraphics[width=8.5cm]{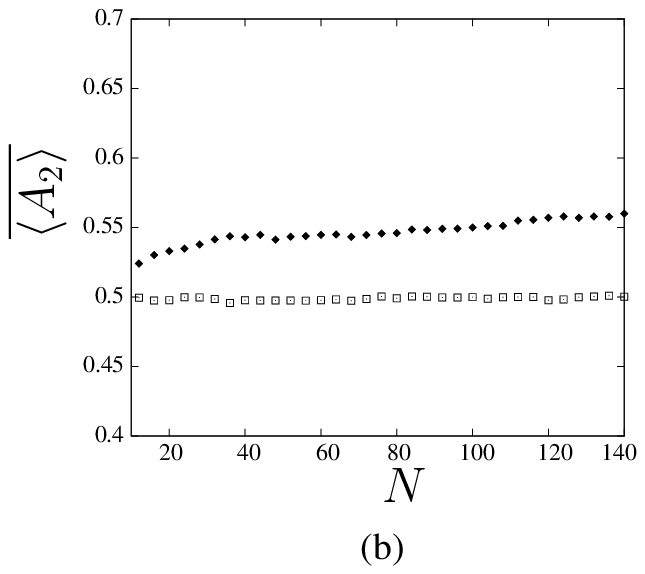}
\caption{ \label{Ad} (a) Averaged three-dimensional asphericity $\overline{\langle A_3 \rangle}$ of a polymer chain 
near a homogeneously attractive 
surface (squares) and an attractive percolation cluster (filled diamonds) as a function of temperature and 
(b) the two-dimensional analog $\overline{\langle A_2 \rangle}$ as a function of $N$ at $T=0.1 \ll T_A$ deep in the adsorbed phase. }\end{center}
\end{figure}

Our results for the averaged asphericity of a polymer grafted to a surface as a function of temperature are given in Fig.~\ref{Ad}(a). 
In the high-temperature regime, both curves tend to the bulk value of $\langle A_3 \rangle$, whereas as temperature decreases, the anisotropy of polymer configuration grows. Note, however, the principal point. When the temperature is well above the adsorption transition, 
the polymer coil in bulk is a three-dimensional object and thus is characterized by the quantity $\langle A_3 \rangle$. However, in the adsorbed state (well below $T_A$), the polymer lies on the surface and can be treated rather like a two-dimensional object, thus $\langle A_2 \rangle$ is the more natural characteristic. The quantity $A_2$ of a two-dimensional object can, however, be simply related to $A_3$, evaluating (\ref{add}), e.g., at $\lambda_3=0$ with arbitrary $\lambda_1$, $\lambda_2$: 
\begin{equation}
A_3 = \frac{1}{4} + \frac{3}{4}A_2. \label{f}
\end{equation} 
Definition (\ref{add}) is an {\em extrinsic\/} measure for the asphericity, depending on
the dimension of the embedding space. Of course, in the present situation, it would be nicer 
to come up with an {\em intrinsic\/} measure, similar to Gauss' curvature definition.
The asphericity $\overline{\langle A_2\rangle}$ of a polymer adsorbed on a homogeneously attractive surface and attractive percolation cluster
is given in Fig.~\ref{Ad}(b) as a function of $N$.
For finite chain length $N$, the values of $\overline{\langle A_2(N)\rangle}$ differ from those for infinitely long chains. This finite-size deviation obeys scaling behavior with $N$:
\begin{eqnarray}
\overline{\langle A_d(N) \rangle}= \langle A_d \rangle +b_1 N^{-\Delta},\label{corr}
\end{eqnarray}
where $b_1$ is a constant and $\Delta$ is the correction-to-scaling exponent: $\Delta(d=2)=1.5$ \cite{Caracciolo05} in the homogeneous case.
The shape parameter estimates can be obtained by least-square fitting of (\ref{corr}).  For the case of the pure lattice, we receive 
$\langle A_2 \rangle=0.502\pm0.006$, whereas for the polymer on the attractive percolation cluster we have 
   $\overline{\langle A_2^{p_c} \rangle}= 0.567\pm0.006$, which within  error bars agrees with the corresponding value found by us previously by 
analyzing the conformational statistics of polymers on underlying percolation clusters \cite{Blavatska10}.
Note, that the corresponding values of the three-dimensional asphericity according to (\ref{f}) are: 
 $\langle A_3 \rangle=0.627\pm0.006$, $\overline{\langle A_3^{p_c} \rangle}= 0.675\pm0.006$, which
 agree with the $T \rightarrow 0$ limit in Fig.~\ref{Ad}(a).
The principal qualitative conclusion from these shape parameters is that 
typical conformations of a polymer chain, which is adsorbed on an attractive  percolation cluster, are more anisotropic than those for a 
homogeneously attractive surface
due to the complicated fractal structure of the adsorbing ``sieve".

\section{Conclusions}

We have studied the adsorption of flexible polymer macromolecules on an attractive 
percolation cluster, formed on a regular two-dimensional disordered lattice at critical concentration $p_c$
of occupied sites.  We treat such disordered surface as 
the $z=0$ plane of a regular simple-cubic three-dimensional lattice. In our model, the sites which do not belong to the percolation cluster,
are penetrable for the polymer chain.
The percolation cluster is a fractal object, characterized by the fractal dimension 
$d_s^{p_c}=91/49\simeq1.89$ {\cite{Havlin87}}, thus we have the problem of polymer adsorption on a
fractal substrate.

The conformational properties of polymer 
chains grafted to a percolation cluster are studied with the pruned-enriched Rosenbluth method (PERM) \cite{Grassberger97}.
%We examined the behavior of components of radius of gyration $\overline{\langle R^2_{g{||}}\rangle}$, 
%$\overline{\langle R^2_{g{\perp}}\rangle}$ in directions parallel and perpendicular to the surface,
%and revealed, that the critical exponent, governing the scaling of the size of the polymer chain adsorbed on a
%fractal substrate  
% formed by a percolation cluster, is smaller than that for a homogeneously attractive surface. A value 
%$\nu_2^{p_c}=0.682\pm0.006$ is obtained, to be compared with  $\nu_2=0.742\pm0.006 \approx 3/4$ for a plain surface.  
%We assume that such a ``shrinking" of the average polymer extension is caused by wrapping of the polymer chain
%around both sides of the percolation cluster. 
Examining the peak structure of the heat capacity, we find an estimate for 
the surface crossover exponent, governing the scaling of the adsorption energy in the vicinity of the transition point,
$\phi_s^{p_c}=0.425\pm0.009$, and for the adsorption transition temperature we obtain $T_A^{p_c}=2.64\pm0.02$.  
As expected, the adsorption is diminished, when the fractal dimension of the surface is 
smaller than that of the plain Euclidean surface due to the smaller number of contacts of monomers
 with attractive sites.

We also analyzed the shape of typical spatial conformations that a polymer chain attains in the adsorbed state.
The asymmetry of the shape can be characterized by rotationally invariant universal quantities, 
such as the asphericity $A_d$, which equals zero for a spherical configuration,  and takes on a maximum value 
of one in the case of a completely stretched rod-like configuration.  For the polymer on the attractive 
percolation cluster we received the value  
   $\overline{\langle A_2^{p_c} \rangle}= 0.567\pm0.006$, which is larger than that on the plain surface, 
$\langle A_2 \rangle=0.502\pm0.006$. The principal qualitative conclusion from our analysis of the
shape parameters is that 
typical conformations of a polymer chain that is adsorbed on an attractive ``sieve'' formed by
a percolation cluster are more anisotropic than those of a homogeneously attractive surface. 

\section*{Acknowledgements}
We thank Niklas Fricke for useful discussions.
Work supported by 
S\"achsische DFG Forschergruppe FOR877 under Grant No.\ Ja 483/29-1,
DFG Sonderforschungs\-bereich SFB/TRR 102 (project B04), and
Graduate School of Excellence GSC 185 ``BuildMoNa''. 
V.B. is grateful for hospitality of the S\"achsische DFG Forschergruppe FOR877 during
an extended research stay in Leipzig.

\end{document}